\documentclass{ws-procs975x65}

\begin{document}

\markboth{Gergely}
{Is dark matter futile on the brane?} 

\title{IS DARK MATTER FUTILE ON THE BRANE? \footnote{
Research supported by OTKA grants no. T046939, TS044665 and the J\'{a}nos
Bolyai Fellowships of the Hungarian Academy of Sciences. The author wishes
to thank the organizers of the 11th Marcel Grossmann Meeting for support.} }
\author{L\'{A}SZL\'{O} \'{A}. GERGELY}

\address{Departments of Theoretical and Experimental Physics, 
University of Szeged,\\
D\'om t\'er 9, H-6720 Szeged, Hungary\\
\email{gergely@physx.u-szeged.hu}}

\begin{abstract}
We investigate whether dark matter can be replaced by various source terms
appearing in the effective Einstein equation, valid on a brane embedded into
a higher dimensional space-time (the bulk). Such non-conventional source
terms include a quadratic (ordinary) matter source term, a geometric source
term originating in the Weyl curvature of the bulk, a source term arising
from the possible asymmetric embedding, and finally the pull-back to the
brane of possible non-standard model bulk fields.
\end{abstract}


\bodymatter

\section*{}

According to modern precision cosmology\cite{SDSSk=0,SDSSWMAPk=0,WMAP3y} our
universe is described by a Friedmann-Lema\^{\i}tre-Robertson-Walker geometry
with flat spatial sections. The matter source for this geometry is however
less well known. Ordinary, baryonic matter gives a tiny $\approx 3\%$
fraction $\rho _{b}$ of the critical energy density. In order to explain
galactic cluster dynamics and galactic rotation curves, dark matter has been
introduced, and it is believed that nowadays there is approximately $10$
times as much dark matter as baryonic matter in the universe. Models of cold
and warm (but not hot) dark matter are compatible with observations. The
remaining $\approx 70\%$ of the critical energy density is given by a
mysterious dark energy, a form of matter represented in the simplest way by
the famous cosmological constant $\Lambda $ (a perfect fluid with pressure $%
p_{\Lambda }=w\rho _{\Lambda }~,$ $w=-1$). Dark energy obeys $\rho
_{de}+3p_{de}<0$, the condition which (due to the Raychaudhuri equation)
yields to an accelerated expansion of the universe. The latter seems
unavoidable in the light of the recent distant supernovae surveys.\cite%
{gold, SNLS} Observations support very well the simplest possible dark
energy, the cosmological constant $\Lambda $ and the basic question is not
how different $w$ is from $-1$, but instead whether it is allowed to have
any dynamics. But even if dark energy would be simply a cosmological
constant, what is its origin? The explanation in terms of vacuum 
energy density turned into one of the biggest problems of modern theoretical
physics: the discrepancy is of 120 orders of magnitude.

With so much unknown matter and energy in our universe the question
naturally arises, whether gravitational dynamics, as we now it, has to be
modified? Dark matter is conveniently replaced by the so-called MOND
gravitational theory.\cite{MOND,MOND2} This however conflicts with
cosmological observations.\cite{MONDconflict} The late-time acceleration 
of the universe can be explained by the inverse
curvature gravity theories.\cite{modgrav}

The theory of gravitation can be also modified by returning to the
pioneering ideas of Kaluza and Klein, allowing for more than $4$ space-time
dimensions. Indeed string / M-theory is multidimensional and in recent years
a simpler class of effective models,\cite{RS1,RS2,MaartensLR} known as
brane-worlds have been intensively studied. In these models the observable
universe is a $4$-dimensional space-time continuum (the brane with tension $%
\lambda $) on which standard model fields act. Gravitation however is
allowed to leak out in a fifth, possibly non-compact dimension. Its $5$%
-dimensional dynamics is governed by the Einstein equation with cosmological
constant term.

The apparent gravitational dynamics on our observable $4$-dimensional
universe appears through a projection formalism. It consist of the twice
contracted Gauss equation, the Codazzi equation and an effective Einstein
equation.\cite{Decomp} In the derivation of the latter the junction
conditions across the brane are employed. The effective Einstein equation,
first derived\cite{SMS} for symmetric embedding and no bulk sources was
later supplemented by the pull-back to the brane of the bulk energy momentum
tensor $\widetilde{\Pi }_{ab}~:$ 
\begin{equation}
\mathcal{P}_{ab}=\frac{2\widetilde{\kappa }^{2}}{3}\left( g_{a}^{c}g_{b}^{d}%
\widetilde{\Pi }_{cd}\right) ^{TF}
\end{equation}%
(with $\widetilde{\kappa }^{2}$ the bulk coupling constant and $g_{ab}$ the
induced metric on the brane) and the asymmetry source term $\overline{L}%
_{ab}^{TF}$ which is the tracefree part of the tensor
\begin{equation}
\overline{L}_{ab}=\overline{K}_{ab}\overline{K}-\overline{K}_{ac}\overline{K}%
_{b}^{c}-\frac{g_{ab}}{2}\left( \overline{K}^{2}-\overline{K}_{ab}\overline{K%
}^{ab}\right) \ ,
\end{equation}%
(with $\overline{K}_{ab}$ the mean extrinsic curvature). The effective
Einstein equation is:\cite{Decomp}%
\begin{equation}
G_{ab}=-\Lambda g_{ab}+\kappa ^{2}T_{ab}+\widetilde{\kappa }^{4}S_{ab}-%
\overline{\mathcal{E}}_{ab}+\overline{L}_{ab}^{TF}+\overline{\mathcal{P}}%
_{ab}\ .  \label{modEgen}
\end{equation}%
(with $\kappa ^{2}$ the brane coupling constant). The function $\Lambda =\left( 
\widetilde{\kappa }^{2}/2\right) (\lambda -n^{c}n^{d}\widetilde{\Pi }_{cd}-%
\overline{L}/4)$ possibly varies due to both the normal
projection of the bulk energy-momentum tensor and the embedding. When
constant, $\Lambda $ represents the the brane cosmological constant. The
source term $S_{ab}$ is quadratic in the brane energy-momentum tensor $T_{ab}
$, and it modifies early cosmology.\cite{BDEL} The quantity 
\[
\mathcal{E}_{ac}=\widetilde{C}_{abcd}n^{b}n^{d}
\]%
represents the electric part of the bulk Weyl tensor $\widetilde{C}_{abcd}$.
In a cosmological context $\mathcal{E}_{ab}$ is known as dark radiation.

Modified gravitational dynamics leads therefore to four new source terms in
the effective Einstein equation. In what follows we discuss the potential of
generating / replacing dark matter, represented by these source terms. 

From
this point of view the less interesting of them is the non-linear source
term $S_{ab}$. This is because it scales as $\rho /\lambda $, as compared to
the energy-momentum tensor and this ratio is infinitesimal, excepting the
very early universe, due to the huge value $\lambda >138.59\,\, $TeV$^{4}$
of the brane tension.\cite{Irradiated}\ 

The Weyl curvature of the bulk on the other hand can generate quite
remarkable electric part contributions $\mathcal{E}_{ab}$. In a spherically
symmetric brane-world metric the Schwarzschild mass parameter receives a new
contribution due to $\mathcal{E}_{ab}$ which is interpreted as the mass of
dark matter.\cite{HarkoCheng}. When the cosmological constant is neglected,
the dark mass scales linearly with the radial distance, explaining the
flatness of the galactic rotation curves. The bending angle of light in such
brane world models was found much larger as compared to the predictions of
the dark matter models, the deviation increasing with the degree of
compactness. This is however exactly the regime where no observational data
on gravitational lensing is available yet.

Properly chosen non-standard model bulk fields can replace dark matter in
explaining structure formation,\cite{Pal} as the evolution of perturbations
on the brane becomes similar to that of the Cold Dark Matter (CDM) model.

Most interestingly, the combined effect of asymmetry and a bulk radiation
qualitatively can give both dark matter and dark energy.\cite%
{Irradiated,Irradiated2} This has been investigated in a closed brane-world
universe model, with bulk black hole only on one side of the brane. In this
model the emitted Hawking radiation is partially absorbed on the brane and
partially transmitted through, but other models with radiation in the bulk
have been also considered.\cite{Jennings} In principle in such models there
are two competing effects: (a) the radiation pressure accelerates the brane,
manifesting itself as dark energy, and (b) the absorbed radiation increases
the energy density of the bulk, appearing as CDM. These two effects compete
with each other, and with properly chosen initial data a critical-like
behavior was observed.\cite{Irradiated,Irradiated2}
Based on the predictions of these toy models, it seems worth to work out 
more realistic brane-world models in order to confront them with observations.

\end{document}